\documentclass[a4paper,11pt]{article}
\usepackage{pos}
\usepackage[ngerman,english]{babel}
\usepackage{longtable,tabularx}
\usepackage{booktabs}
\usepackage{aastex_hack}

\usepackage{graphicx}
\usepackage{wrapfig}

\title{Probing Galaxy structure with VHE $\gamma$ rays}

\author*[a]{Constantin Steppa}
\author[a]{Kathrin Egberts}

\affiliation[a]{Institut f\"ur Physik und Astronomie, Universit\"at Potsdam,\\
  Karl-Liebknecht-Str. 24, 14476 Potsdam, Germany}

\emailAdd{steppa@uni-potsdam.de}
\emailAdd{kathrin.egberts@uni-potsdam.de}

\abstract{As an observer from within the Milky Way, it is difficult to determine its global structure. Despite extensive observational data from surveys at different wavelengths, we have no conclusive description of the structure of our own Galaxy. For very-high-energy (VHE) $\gamma$ rays, the most comprehensive catalogue of Galactic sources resulting from the H.E.S.S. Galactic Plane Survey (HGPS) shows a striking asymmetry in the distribution of the sources in the latitudinal direction. This could be the result of a local feature in the spatial distribution of the sources or it could be due to the position of the Sun above the Galactic plane. In this contribution, we estimate the position of the Sun based on the latitudinal flux profile of VHE $\gamma$-ray sources, assuming three mirror-symmetric models for the spatial distribution of the sources in three-dimensional space and taking into account the observational bias of the HGPS. We verify our method using simulations and find values for $z_{\odot}$ between $-6\,\mathrm{pc}$ and $94\,\mathrm{pc}$ depending on the considered model. Our results show that the position of the Sun has a significant impact on the observed source distribution and must therefore be taken into account when modelling the population of Galactic VHE $\gamma$ sources. However, it is not conclusive whether the Sun's offset from the Galactic plane is the only factor leading to the asymmetry in the latitudinal profile.
}

\FullConference{%
  7th Heidelberg International Symposium on High-Energy Gamma-Ray Astronomy (Gamma2022)\\
  4-8 July 2022\\
  Barcelona, Spain\\}

\begin{document}
\maketitle

\section{Introduction}
\begin{wrapfigure}{O}{.5\textwidth}
  \includegraphics[width=0.5\textwidth]{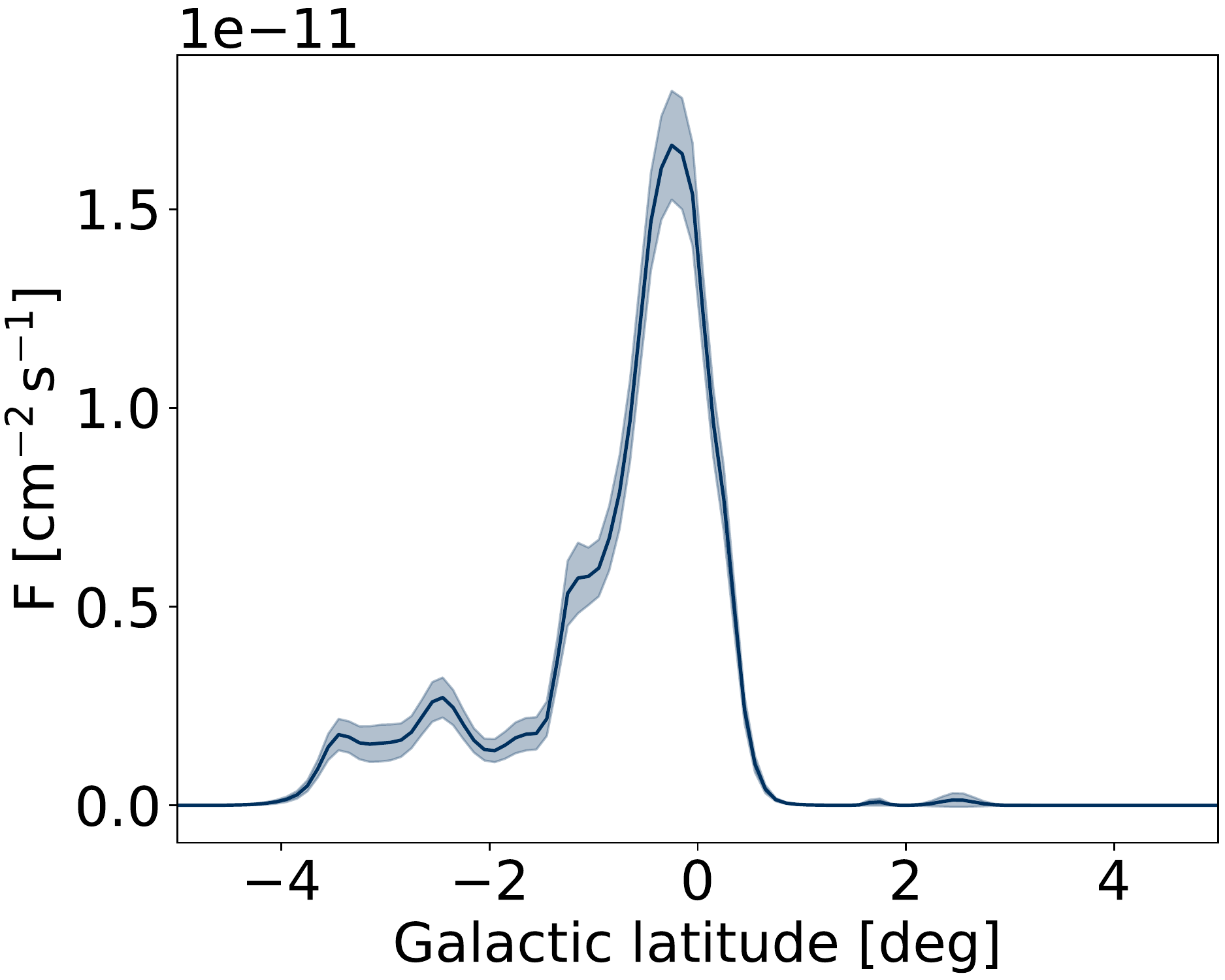}
  \caption[Latitudinal profile of HGPS sources]{Flux profile along Galactic latitude of sources in the H.E.S.S. Galactic plane survey.}\label{FIG:profile}
\end{wrapfigure}
In addition to our unfavourable observational position in the Galactic plane, at very high energies (VHE) it is not possible to model the spatial distribution of sources based on the observational data alone, due to the small number of known $\gamma$-ray sources and the lack of information on their distance from us. Alternatively, however, we can use observations at VHE in combination with distribution models obtained, for example, on the basis of more extensive observations of typical $\gamma$-ray source classes in other wavelength ranges to derive the properties of the total population of Galactic VHE $\gamma$-ray sources. As shown in \cite{Steppa2020}, the choice of the spatial-distribution model has a strong influence on the derived properties of the population, for example on the total number of sources and their luminosities. This means that the most accurate possible description of the spatial source distribution is crucial to make reliable predictions derived from the model, such as the contribution of the unresolved sources to the measured diffuse emission.\\
This work focuses on an aspect where all models investigated in \cite{Steppa2020} have difficulties in describing the observational data, namely the asymmetry apparent in the latitudinal profile of the known $\gamma$-ray sources shown in Fig.~\ref{FIG:profile}. In particular, we investigate whether this observation can be explained by the position of the Sun above the Galactic plane. An offset of the Sun above the plane would cause a compression of the distribution of observed sources in the northern direction and a stretching of the distribution in the southern direction. For this study, we utilise the source catalogue of the H.E.S.S. Galactic Plane survey (HGPS) \cite{Hess2018} and use the three most promising population models from \cite{Steppa2020}, namely mPWN, mSNR and mSp4. The spatial source distribution of the first model is based on the observation of pulsars \cite{Yusifov2004, Lorimer2006}, for the second on that of supernova remnants \cite{Green2015,Xu2005}. Both models are axially symmetric. The third is a spiral-arm model and is based on the measurement of interstellar matter \cite{SteimanCameron2010}. All models are mirror-symmetrical with respect to the Galactic plane. The distributions of the luminosities and physical extensions of the sources are each described with a simple power law. The exact parameterisation of the models is given in \cite{Steppa2020}. In the following, we describe the method to determine the most likely position of the Sun, which best describes the data for a given model, and discuss the results.

\section{Method}
\begin{figure}
  \centering
  \includegraphics[width=.95\textwidth]{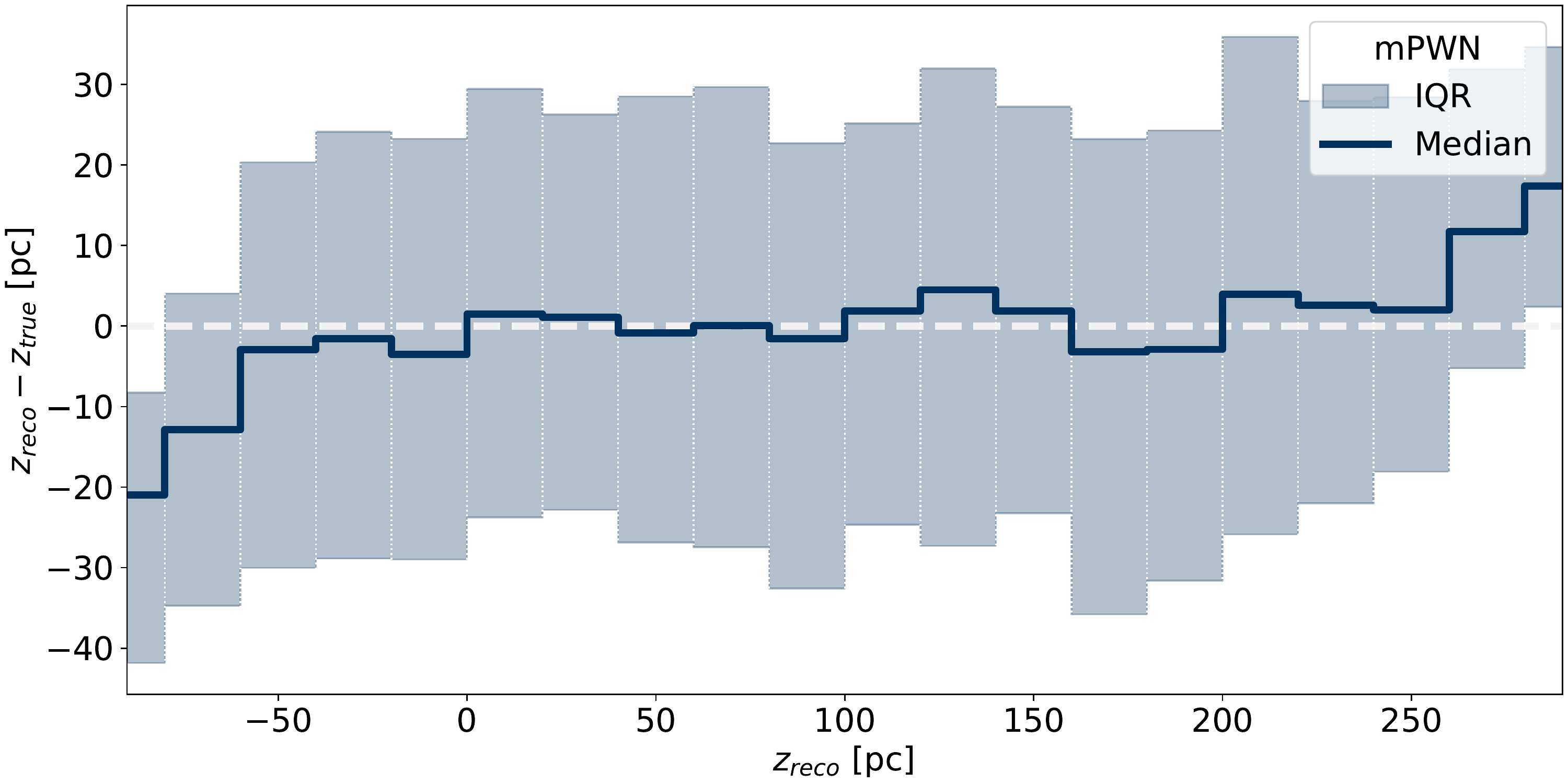}
  \caption[Verification of the method]{Distribution of the error on the reconstructed position of the Sun. It was derived from simulations of $10^4$ source populations based on the model \textit{mPWN}. Depending on the reconstructed position $z_{reco}$, the solid line shows the median of the distribution of the error ($z_{reco} - z_{true}$) over all populations yielding the respective $z_{reco}$ value. Likewise the shaded area shows the interquartile range of the error distribution respectively.}
	\label{FIG:validation}
\end{figure}

For each model, we simulated $10^{4}$ populations of sources that are characterised by the sources' spatial positions, luminosities and physical extensions. For these synthetic populations we then calculated their observational properties, i.e. position on the sky, flux and angular extent, depending on the assumed position of the Sun above the plane $z_{\odot}$. From that the latitudinal flux profile is derived for sources that would be detectable by the HGPS, taking into account its observation bias. For the fluxes, we calculate errors by assuming a systematic error of $30\,\%$ on the source flux as given in the HGPS \cite{Hess2018}. If we compare the HGPS data with a synthetic population at latitude bin $i$, then the probability that we measure the HGPS flux $F_{HGPS,i}$---assuming that the synthetic population gives us the true value---results from a Gaussian distribution $\mathcal{N}(F\, |\,\mu_{i}, \sigma_{i})$ whose mean $\mu_{i}$ is given by the flux of the synthetic population at bin $i$ and the width $\sigma_{i}$ is given by the error on the flux of the synthetic population. However, since the synthetic population is only one realisation of our model, the next step is to average the probability over $N$ realisations to obtain the probability of the HGPS data under our model assumption
\begin{equation}
  Prob(F_{HGPS,i}\,|\,z_{\odot}) = \frac{1}{N}\sum_{j=1}^{N}\mathcal{N}(F_{HGPS, i}\,|\,\mu_{i,j}, \sigma_{i,j}, z_{\odot})\,.
\end{equation}
A likelihood for the observation is then calculated from all $M$ bins in the profile as
\begin{equation}
  \mathcal{L}(z_{\odot}) = \prod_{i=1}^{M}Prob(F_{HGPS, i}\,|\,z_{\odot})\,.
  \label{EQ:likelihood}
\end{equation}
Equation \ref{EQ:likelihood} allows us to study the effect of the position of the Sun on the shape of the latitudinal profile on a statistical basis by varying $z_{\odot}$ in the simulations. The most likely position, i.e. the one that describes the data best according to the considered model, is determined by the maximum likelihood. Since our calculation of the likelihood is based on Monte Carlo simulations, it is subject to statistical fluctuations. Tests have shown that the choice of the synthetic populations used for the calculation can have a considerable influence on the result. This behaviour ultimately contributes to the uncertainty in the determination of $z_{\odot}$. To make the result of our estimation more robust, we divided the $10^{4}$ synthetic populations into sets of 200 populations each and calculated the likelihood for each set. We then used the median of the likelihood over these 50 sets to determine $z_{\odot}$.\\
We verified our method of estimating $z_{\odot}$ on simulations by applying the method to independent test sets of $10^4$ simulated populations per spatial model with randomly chosen $z_{\odot}$ in the range $[-100,\,300]\,\mathrm{pc}$. Exemplarily, the difference between the reconstructed and the simulated position of the Sun ($z_{reco} - z_{true}$) (henceforth error) as a function of the reconstructed position $z_{reco}$ is shown in Fig.~\ref{FIG:validation} for the model \textit{mPWN}.
There it can be seen that the error distribution is generally well centred around $0\,\mathrm{pc}$, which attests to the method's good accuracy. However, the precision of this method is rather poor, with a typical spread in the error distribution $>20\,\mathrm{pc}$. Important factors leading to this large spread are the small number of detected sources and missing information regarding the distances to the sources. The error distribution is similar for the other two models.

\section{Result}
\begin{figure}
  \centering
  \includegraphics[width=.95\textwidth]{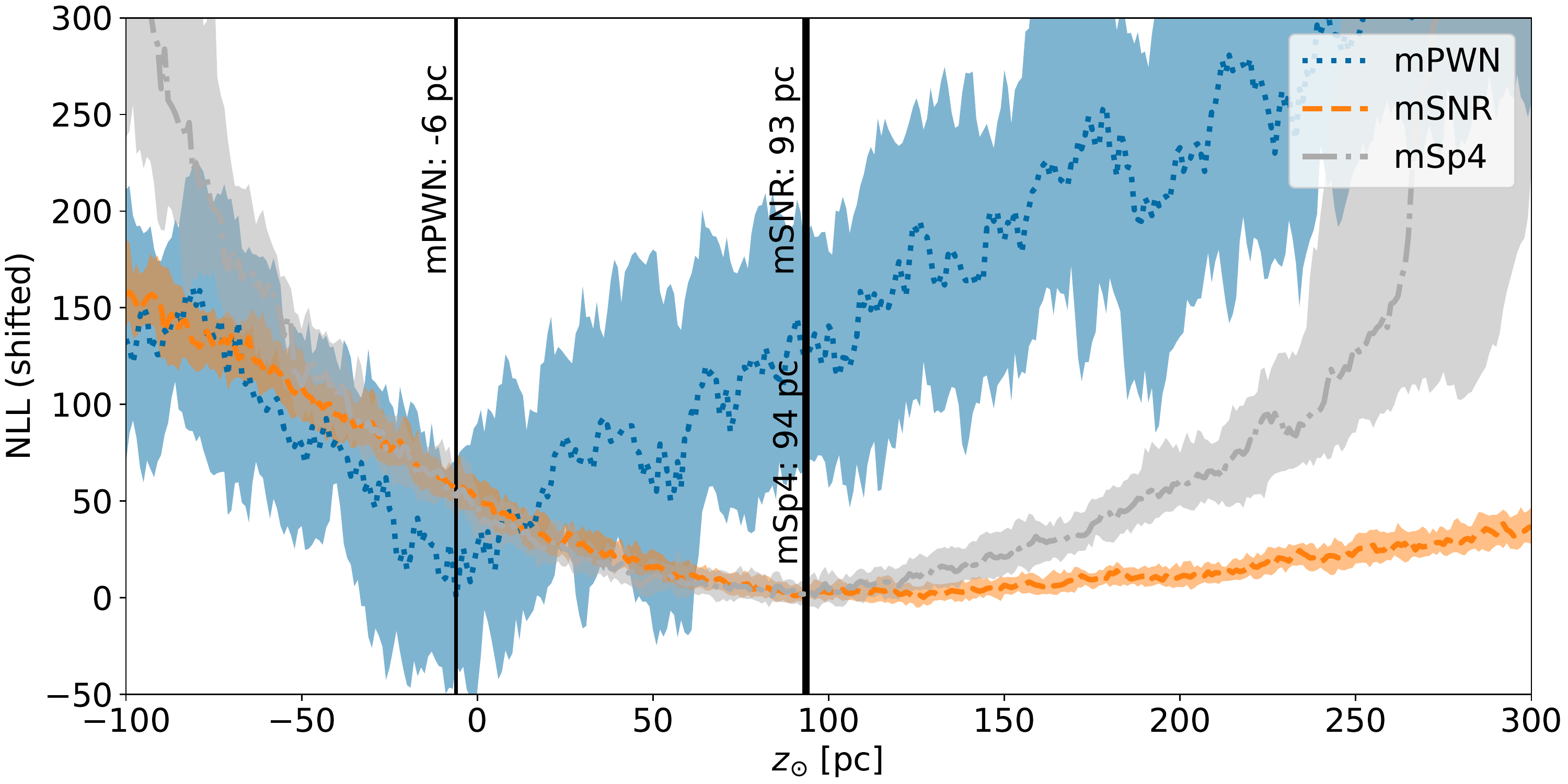}
  \caption[Negative log-likelihood]{The negative log-likelihood (NLL) as a function of $z_{\odot}$ for the different spatial models. The lines show the median of the NLL calculated from 50 sets of 200 synthetic populations each. The shaded areas show the interquartile range accordingly. The results shown here are shifted so that for each model the minimum of the median distribution is zero. Vertical lines mark the minima of the different spatial models.}
	\label{FIG:nll}
\end{figure}

To determine the optimal position of the sun, we calculated the likelihood for $z_{\odot}$ in the range $[-100, 300]\,\mathrm{pc}$ with a resolution of $1\,\mathrm{pc}$. The negative log-likelihood (NLL) is shown in Fig.~\ref{FIG:nll}, with the result for the different models shifted vertically for better visualisation.
The lines show the median NLL calculated from 50 sets of 200 synthetic populations each. The shaded areas show the interquartile range accordingly. The model mPWN shows strong variation in the NLL distribution, but with a clear minimum near $0\,\mathrm{pc}$. In comparison, the model mSp4 seems to be less affected by variations in the calculated NLL. It yields a minimum at $z_{\odot}=94\,\mathrm{pc}$. This value for $z_{\odot}$ is significantly larger than that for mPWN, whereas the minimum is less pronounced. Similarly, the model mSNR yields a minimum at almost the same value as mSp4, namely at $z_{\odot}=93\,\mathrm{pc}$. In contrast to the other models, we find a flat distribution of the NLL towards larger $z_{\odot}$ for mSNR. All values are listed in Tab.~\ref{TAB:result} together with their corresponding errors. The errors result from the interquartile range of the error distribution, as shown for example in Fig.~\ref{FIG:validation}.

\begin{longtable}[c]{lr}
  \caption[Estimates of $z_{\odot}$]{Estimates of $z_{\odot}$.}
  \label{TAB:result} \\
  \toprule
  Model & $z_{\odot}$ $\mathrm{[pc]}$ \\
  \midrule
  \endfirsthead
  \bottomrule
  \endfoot
  mPWN & $-6_{-25}^{+27}$ \\
  mSNR & $93_{-70}^{+47}$ \\
  mSp4 & $94_{-44}^{+40}$ \\
\end{longtable}

\section{Discussion}
The results show a twofold picture. On the one hand, there are the models mSNR and mSp4, for which a position of the Sun at about $90\,\mathrm{pc}$ above the Galactic plane best reflects the observed latitudinal profile and its inherent asymmetry. These models are characterised by a thin Galactic disk and correspondingly a narrow latitudinal profile, as observed for the HGPS. A large value of $z_{\odot}$ is required in these models to stretch the profile of the synthetic source populations to match the observed profile. For both models, a likelihood ratio test shows that the solar offset hypothesis is clearly preferable to the null hypothesis with the Sun position in the Galactic plane (p-values $<10^{-22}$). On the other hand, a thicker Galactic disk is assumed for the mPWN model, resulting in a broad latitudinal profile. Although the position of the Sun has a clear impact on the latitudinal profile derived with this model, as we observe from the distribution of NLL (see Fig.~\ref{FIG:nll}), the obtained value for $z_{\odot}$ is compatible with $0\,\mathrm{pc}$, taking into account its errors. It appears that with this model it is more likely that the observed asymmetry is nothing more than a product of statistical fluctuations in a model realisation. This means the asymmetry would be the result of a local feature in the source distribution. However, the steep drop in the HGPS latitudinal profile towards high Galactic latitude generally contrasts with the broad profile of this model. At this point we would like to point out that for the present study we have resorted to models whose optimisation of the model parameters for the distribution of source luminosities and extents is based on the assumption of an in-plane solar position. It is quite possible that a different assumption would also affect these model parameters and thus the latitudinal profile.\\
The new measurement of the Sun's Galactic height that we have presented in this work is based exclusively on VHE $\gamma$-ray data. Due to the resolution of current instruments for measuring VHE $\gamma$-rays, the amount and information content of these data is limited. Therefore, a comparison with results obtained in other wavelength ranges and with different methods is instructive. Typically, those measurements can rely on more extensive and precise data sets. Recent measurements, based for example on BeSSel or Gaia data, place the Sun at an height of $\sim 5-20\,\mathrm{pc}$ \cite{Reid2019,Bennett2019}. The relatively wide range of values obtained for this quantity reveals the difficulty of its measurement and a dependence of the values obtained on the data and models consulted. With respect to the values obtained in this work, only the model mPWN shows compatibility within errors with the cited measurements. However, the models mSNR and mSp4 seem to qualitatively provide a better description of the VHE data due to the width of the latitudinal profile. With models that describe the VHE data better but are incompatible with other measurements on the one hand, and a model where the opposite is the case on the other hand, it is not possible to make a final judgement as to whether the asymmetry in the latitudinal profile is a product of the position of the Sun or a local feature in the source distribution. Ultimately, there may be more than one factor that play a role here. Overall, however, we were able to confirm with this work that the position of the Sun above the Galactic plane, as indicated by a large number of measurements, also affects the population of Galactic $\gamma$-ray sources and must therefore be taken into account in its modelling.\\
An extension of the source sample, for example by the Cherenkov Telescope Array \cite{CTA2019}, and extensive multi-wavelength studies that allow the identification of sources and, in particular, distance determination, will help to better understand the connection between the global structure of the Milky Way and the sources of the VHE $\gamma$ rays in the near future.

\bibliographystyle{JHEP}
\bibliography{galaxy_structure}

\providecommand{\cjaa}{Chinese Journal of Astronomy~\& Astrophysics}

\providecommand{\href}[2]{#2}\begingroup\raggedright\begin{thebibliography}{10}

\bibitem{Steppa2020}
C.~{Steppa} and K.~{Egberts}, \emph{{Modelling the Galactic very-high-energy
  {\ensuremath{\gamma}}-ray source population}},
  \href{https://doi.org/10.1051/0004-6361/202038172}{\emph{\aap} (2020) }.

\bibitem{Hess2018}
{H.~E.~S.~S. Collaboration}, H.~{Abdalla}, A.~{Abramowski}, F.~{Aharonian},
  F.~{Ait Benkhali}, E.O.~{Ang{\"u}ner} et~al., \emph{{The H.E.S.S. Galactic
  plane survey}},
  \href{https://doi.org/10.1051/0004-6361/201732098}{\emph{\aap} (2018) }.

\bibitem{Yusifov2004}
I.~{Yusifov} and I.~{K{\"u}{\c{c}}{\"u}k}, \emph{{Revisiting the radial
  distribution of pulsars in the Galaxy}},
  \href{https://doi.org/10.1051/0004-6361:20040152}{\emph{\aap} (2004) }.

\bibitem{Lorimer2006}
D.R.~{Lorimer}, A.J.~{Faulkner}, A.G.~{Lyne}, R.N.~{Manchester}, M.~{Kramer},
  M.A.~{McLaughlin} et~al., \emph{{The Parkes Multibeam Pulsar Survey - VI.
  Discovery and timing of 142 pulsars and a Galactic population analysis}},
  \href{https://doi.org/10.1111/j.1365-2966.2006.10887.x}{\emph{\mnras} (2006)
  }.

\bibitem{Green2015}
D.A.~{Green}, \emph{{Constraints on the distribution of supernova remnants with
  Galactocentric radius}},
  \href{https://doi.org/10.1093/mnras/stv1885}{\emph{\mnras} (2015) }.

\bibitem{Xu2005}
J.-W.~{Xu}, X.-Z.~{Zhang} and J.-L.~{Han}, \emph{{Statistics of Galactic
  Supernova Remnants}},
  \href{https://doi.org/10.1088/1009-9271/5/2/007}{\emph{\cjaa} (2005) }.

\bibitem{SteimanCameron2010}
T.Y.~{Steiman-Cameron}, M.~{Wolfire} and D.~{Hollenbach}, \emph{{COBE and the
  Galactic Interstellar Medium: Geometry of the Spiral Arms from FIR Cooling
  Lines}}, \href{https://doi.org/10.1088/0004-637X/722/2/1460}{\emph{\apj}
  (2010) }.

\bibitem{Reid2019}
M.J.~{Reid}, K.M.~{Menten}, A.~{Brunthaler}, X.W.~{Zheng}, T.M.~{Dame}, Y.~{Xu}
  et~al., \emph{{Trigonometric Parallaxes of High-mass Star-forming Regions:
  Our View of the Milky Way}},
  \href{https://doi.org/10.3847/1538-4357/ab4a11}{\emph{\apj} (2019) }.

\bibitem{Bennett2019}
M.~{Bennett} and J.~{Bovy}, \emph{{Vertical waves in the solar neighbourhood in
  Gaia DR2}}, \href{https://doi.org/10.1093/mnras/sty2813}{\emph{\mnras} (2019)
  }.

\bibitem{CTA2019}
{Cherenkov Telescope Array Consortium}, B.S.~{Acharya}, I.~{Agudo}, I.~{Al
  Samarai}, R.~{Alfaro}, J.~{Alfaro} et~al., \emph{{Science with the Cherenkov
  Telescope Array}} (2019),
  \href{https://doi.org/10.1142/10986}{10.1142/10986}.

\end{thebibliography}\endgroup

\end{document}